\documentclass[pra,notitlepage,twocolumn,amsmath,amssymb,aps,superscriptaddress]{revtex4-2}

\usepackage{amsmath,amssymb,amsfonts,bbm,times}
\usepackage[dvipsnames]{xcolor}
\usepackage[T1]{fontenc}
\usepackage{braket}
\usepackage[normalem]{ulem}
\usepackage[colorlinks=true,bookmarks=false,linkcolor=RoyalBlue,urlcolor=RoyalBlue,citecolor=RoyalBlue,breaklinks]{hyperref}
\usepackage{graphicx}
\usepackage{bm}
\usepackage{physics}

\usepackage{soul}

\begin{document}

\title{Gaussian boson sampling at finite temperature}

\author{Gabriele Bressanini}
\affiliation{QOLS, Blackett Laboratory, Imperial College London, London SW7 2AZ, United Kingdom}
\author{Hyukjoon Kwon}
\affiliation{Korea Institute for Advanced Study, Seoul 02455, South Korea}
\author{M.S. Kim}
\affiliation{QOLS, Blackett Laboratory, Imperial College London, London SW7 2AZ, United Kingdom}
\affiliation{Korea Institute for Advanced Study, Seoul 02455, South Korea}

\begin{abstract}
Gaussian boson sampling (GBS) is a promising candidate for an experimental demonstration of quantum advantage using photons.
However, sufficiently large noise might hinder a GBS implementation from entering the regime where quantum speedup is achievable.
Here, we investigate how thermal noise affects the classical intractability of generic quantum optical sampling experiments, GBS being a particular instance of the latter.
We do so by establishing sufficient conditions for an efficient simulation to be feasible, expressed in the form of inequalities between the relevant parameters that characterize the system and its imperfections.
We demonstrate that the addition of thermal noise 
$-$ modeled by (passive) linear optical interaction between the system and a Markovian thermal bath $-$
has the effect of tightening the constraints on the remaining noise parameters, required to show quantum advantage.
Furthermore, we show that there exists a threshold temperature, under the assumption of a uniform loss rate, at which quantum sampling experiments become classically simulable, and provide an intuitive physical interpretation by relating this occurrence with the disappearance of the quantum state's non-classical properties.
\end{abstract}
\maketitle

\section{Introduction}
Gaussian boson sampling (GBS) is a computational task that, under widely accepted complexity-theoretic conjectures, is believed to be intractable using classical machines \cite{gbsog,detailed_gbs,grier2022complexity}. 
The original formulation of the problem consists of sampling from a non-classical $m-$mode Gaussian state obtained by sending identical squeezed vacuum states through a passive linear optical network (LON), with photon-number-resolving (PNR) detectors. 
More experimentally-feasible variants of the task employing threshold detectors \cite{gbs_threshold} and click-counting detectors \cite{gbs_click} have been proposed since.
The advancements of photonic platforms in recent years made an experimental GBS demonstration of quantum advantage feasible with current technological capabilities, with state-of-the-art experiments consisting of 216 modes and up to 125 measured photons \cite{xanadu_experiment}.
Besides the quest for quantum advantage, Gaussian boson samplers can also be used to tackle problems of practical interest such as simulating molecular vibronic spectra \cite{vibronic}, measuring graph similarity \cite{graph_similarity}, perfect matching counting \cite{perfect_matchings}, identifying dense subgraphs \cite{arrazola2018using,graph_problems_chinese}, and predicting stable molecular docking configurations for drug design \cite{banchi2020molecular,yu2022universal}.

As near term quantum devices do not benefit from error correction, it is well known that sufficient noise can prevent GBS experiments from entering the regime where quantum advantage is, in principle, attainable.
Extensive research has been conducted to investigate the impact of various sources of noise and imperfections on the classical simulability of the sampling task.
These include losses \cite{liu2023complexity}, partial distinguishability \cite{shi2022effect,partial_dist}, and detectors' inefficiencies and random counts \cite{RahimiKeshari,classical_simulation_noisy_GBS}.
In particular, Ref.~\cite{RahimiKeshari} provides sufficient conditions for efficient classical simulability of generic quantum optical experiments $-$ GBS being a specific instance $-$ expressed in the form of inequalities that involve the noise parameters.
The method is based on expressing the output probability distribution in terms of phase-space quasi-probability distributions (PQDs) of the input state, the measurement operators and the transition function associated with the specific quantum process, and further identifies their negativity as a necessary condition to achieve quantum advantage \cite{eisert_negativity,veitch2012negative}.

It is well known that a sampling task with thermal state inputs can be efficiently simulated (i.e.,  in polynomial time) on a classical computer \cite{thermal_easy}. 
This fact suggests that there should exist a transition in the computational complexity of GBS as temperature grows.
Nevertheless, finite-temperature effects have received limited attention in this setting thus far.
This paper investigates in this direction by assessing how the addition of thermal noise affects the classical intractability of simulating quantum optical sampling experiments.
We model thermal noise by introducing additional environmental modes initialized in thermal states that interact with the system's modes via a (passive) linear bosonic channel \cite{markovian_noise}.
Significant attention is then dedicated to predicting how much thermal noise can a GBS experiment tolerate before it becomes classically efficiently simulable.

In addition to fundamental interest, thermal noise effects are particularly relevant for experiments conducted in the MHz domain, e.g. those involving the phononic modes of motion of trapped ions \cite{trapped_ions_1,trapped_ions_2}, or those in the GHz domain that employ superconducting architectures.
These platforms offer highly efficient photon-number-resolving detection \cite{pnrd}, enabled by quantum-non-demolition measurements \cite{qnd} that allow for repeated detection, ultimately leading to higher measurement fidelities.
Moreover, superconducting circuits provide an excellent degree of control over the required interactions $-$ namely beam splitter and phase shifter operations $-$ to build an arbitrary passive LON \cite{microwave_BS}. 
Squeezing and displacement operations may also be achieved in circuit QED, thus allowing for the implementation of GBS experiments \cite{GBS_superconducting1}.
The non-linearities provided by Josephson junctions enable efficient preparation of non-classical states of light \cite{synthesizing}, including multi-photon Fock states \cite{fock_states_superconducting,hofheinz2008generation}, and make it possible to engineer non-linear operations that would otherwise be challenging to implement in optical systems. A notable example is given by Kerr-type unitaries, which have recently been introduced into the boson sampling framework as a mean to enhance the system’s robustness against noise \cite{nonlinear_BS}.

The rest of this paper is structured as follows. In Sec.~\ref{sec_sufficientconditions} we revise key aspects of the phase-space formulation of quantum optics and generalize the formalism introduced in Ref.~\cite{RahimiKeshari} by including finite-temperature effects. This allows us to obtain a general sufficient condition for the classical simulability of a generic quantum optical experiment, that we then apply to a noisy GBS instance employing threshold detectors. We find that, as one might expect, the additional thermal noise has the effect of reducing the detection imperfections sufficient to efficiently simulate the sampling task. 
In Sec.~\ref{sec_thermal_classicalization} we show that, under the assumption of a uniform loss rate, there exists a threshold temperature $-$ which depends on the system's losses $-$ at which any sampling experiment becomes classically simulable, even in the presence of ideal detection. We provide a physical interpretation of this phenomenon and show that it is linked to the disappearance of the  state's genuine quantum properties.
In Sec.~\ref{sec_approximate_sampling} we build on the approach of Ref.~\cite{classical_simulation_noisy_GBS} and present a sufficient condition for the classical simulability of noisy GBS experiments that takes into account both thermal noise and approximate sampling, overcoming the main limitation of Ref.~\cite{RahimiKeshari}. 
Lastly, in Sec.~\ref{sec_conclusions} we summarize our findings and provide concluding remarks.

\section{Sufficient conditions for efficient classical simulation of quantum optics at finite temperature}
\label{sec_sufficientconditions}

In this section we investigate how thermal noise affects the noise thresholds in Ref.~\cite{RahimiKeshari} that are sufficient for an efficient classical simulation of a generic bosonic experiment.
Our main result resides in Eq.~\eqref{classicality_condition}, the only assumption to derive such classical simulability condition being the model of noisy evolution we employ, outlined later in this section.
We then consider the special case of a noisy GBS device and show that the latter may be efficiently simulated by classical means if the following inequality is satisfied
\begin{equation}
    \frac{p_D}{\eta_D}\geq \frac{\eta_L}{2}(1-e^{-2r}) - \overline{n}(1-\eta_L)\, .
\end{equation}
Here, $r$ is the squeezing parameter of the input states, $\overline{n}$ is the mean number of environmental thermal photons, $\eta_L$ denotes the transmission of the LON, and $\eta_D$ and $p_D$ are the threshold detectors' efficiency and their dark count rate, respectively.

A generic quantum optical experiment is described in terms of an $m$-mode initial state $\rho$, an $m$-mode quantum process represented by a completely positive (CP) map $\mathcal{E}$ and a measurement performed on the final state $\mathcal{E}(\rho)$.
A quantum measurement is characterized by a positive operator-valued measure (POVM), i.e., a collection of operators  $\lbrace \Pi_{\bm{n}} \rbrace$ satisfying the conditions $\Pi_{\bm{n}}\geq 0$ and $\sum_{\bm{n}}\Pi_{\bm{n}}=\mathcal{I}$, where $\mathcal{I}$ denotes the identity operator on the Hilbert space.
The probability of obtaining a specific measurement outcome $\bm{n}$ is given by the Born rule $p(\bm{n})=\Tr{\mathcal{E}(\rho)\Pi_{\bm{n}}}$. The outcome probability can alternatively be expressed in terms of ordered phase-space quasi-probability distributions as
\begin{equation}
    p(\bm{n})=\pi^m\!\int\! d^{2m}\bm{\beta}\!\! \int \! d^{2m}\bm{\alpha} \,  W_{\Pi_{\bm{n}}}^{(-\bm{s})}(\bm{\beta}) T_{\mathcal{E}}^{(\bm{s},\bm{t})}(\bm{\alpha},\bm{\beta}) W_{\rho}^{(\bm{t})}(\bm{\alpha}) \, .
    \label{born_rule_in_phase_space}
\end{equation}
Here, $W_{\Pi_{\bm{n}}}^{(-\bm{s})}$ and $W_{\rho}^{(\bm{t})}$ denote the PQD of the POVM element $\Pi_{\bm{n}}$ and that of the input state $\rho$, respectively.
In what follows, the dagger transposes a vector of complex numbers to a column vector and takes a complex conjugate.
The $\bm{s}-$ordered PQD ($\bm{s}-$PQD) of a generic $m$-mode Hermitian operator $O$ is defined as
\begin{equation}
    W_{O}^{(\bm{s})}(\bm{\beta}) = \int \frac{d^{2m}\bm{\xi}}{\pi^{2m}} \,
    \Tr{O D(\bm{\xi})}e^{\frac{\bm{\xi} \bm{s} \bm{\xi}^\dagger}{2}}
     e^{\bm{\beta}\bm{\xi}^\dagger-\bm{\xi}\bm{\beta}^\dagger} \, .
     \label{spqd_definition}
\end{equation}
Here, $\bm{s}=\text{diag}(s_1,\dots,s_m)$ is the diagonal matrix of the ordering parameters $s_j \in\mathbb{R}$ and $D(\bm{\xi})$ is the $m$-mode displacement operator 
\begin{equation}
D(\bm{\xi})=e^{\bm{\xi}\bm{a}^\dagger-\bm{a}\bm{\xi}^\dagger} \, ,
\end{equation}
$\bm{a}=(a_1,\dots,a_m)$ being the vector of bosonic operators.
The well known Husimi $Q$-function, Wigner function and Glauber-Sudarshan $P$-function are retrieved for $\bm{s}=-\mathbb{I}_m$, $\bm{s}=0$ and $\bm{s}=\mathbb{I}_m$ respectively, where $\mathbb{I}_m$ denotes the $m-$dimensional identity matrix.
It is worth noting that the $\bm{s}-$PQD of a quantum state is normalized to one, however it can in general attain negative values and diverge more severely than a delta function.
The remaining function appearing in Eq.~\eqref{born_rule_in_phase_space} is the transition function associated with the quantum process $\mathcal{E}$, defined as
\begin{equation}
\begin{split}
    T_{\mathcal{E}}^{(\bm{s},\bm{t})}(\bm{\alpha},\bm{\beta}) & =  \int \frac{d^{2m}\bm{\zeta}}{\pi^{2m}} e^{\frac{\bm{\zeta} \bm{s} \bm{\zeta}^\dagger}{2}}e^{\bm{\beta}\bm{\zeta}^\dagger-\bm{\zeta}\bm{\beta}^\dagger}\int \frac{d^{2m}\bm{\xi}}{\pi^{2m}} e^{-\frac{\bm{\xi} \bm{t} \bm{\xi}^\dagger}{2}}\\ & e^{\bm{\xi}\bm{\alpha}^\dagger-\bm{\alpha}\bm{\xi}^\dagger} \,  \Tr{\mathcal{E}(D^\dagger (\bm{\xi}))D(\bm{\zeta})} \, .
    \label{transition_function}
\end{split}
\end{equation}
One can also prove that
\begin{equation}
    \mathcal{E}(D^\dagger(\bm{\xi}))=e^{\frac{\bm{\xi \xi}^\dagger}{2}} 
    \int \frac{d^{2m}\bm{\gamma}}{\pi^{m}} 
    e^{\bm{\gamma \xi}^\dagger - \bm{\xi \gamma}^\dagger} \mathcal{E}({\ketbra{\bm{\gamma}}}) \, ,
    \label{expansion}
\end{equation}
meaning that the action of $\mathcal{E}$ on a multimode coherent state $\ket{\bm{\gamma}}$ is enough to completely characterize the transition function in Eq.~\eqref{transition_function}.
If there exist values of the ordering parameters $\bm{t}$ and $\bm{s}$ such that $W_{\Pi_{\bm{n}}}^{(-\bm{s})}(\bm{\beta}), \, T_{\mathcal{E}}^{(\bm{s},\bm{t})}(\bm{\alpha},\bm{\beta})$ and $W_{\rho}^{(\bm{t})}(\bm{\alpha})$ are all non-negative and well-behaved, then it is possible to sample from $p(\bm{n})$ efficiently.
We emphasize that this condition is only sufficient, and there might exist other efficient simulation strategies that succeed even in regimes where the PQDs exhibit negativities.
It should also be noted that this framework lets us address the feasibility of efficient \emph{exact} simulations only, i.e., sampling from $p(\bm{n})$, rather than sampling from an approximation of the latter.
Despite this shortcoming, the strength of this formalism resides in the wide range of applicability enabled by its modular nature, which allows us to investigate the classical simulability of generic quantum optical experiments.

Let us now assume that $\mathcal{E}$ is a CP map  describing a LON subject to both photon loss and thermal noise. We adopt a simple model where, alongside the system's $m$ modes, we consider $m$ additional environmental modes, each initialized in the thermal state
\begin{equation}
    \nu_{th}= \frac{1}{1+\overline{n}} \left(\frac{\overline{n}}{1+\overline{n}}\right)^{a^\dagger a} \, .
    \label{thermal_state_def}
\end{equation}
Here $\overline{n}$ is the mean photon number for the given temperature and $a$ is the annihilation operator of the corresponding environmental mode.
These $2m$ modes interact by means of a fictitious ideal interferometer described by the block unitary matrix 
\begin{equation}
    \bm{U}=\begin{pmatrix}
    \bm{L} & \bm{N} \\ \bm{P} & \bm{M} 
    \end{pmatrix} \, .
\end{equation}
The unitarity of $\bm{U}$ implies that 
\begin{equation}
    \bm{L}^\dagger \bm{L} + \bm{P}^\dagger \bm{P} = \mathbb{I}_m \, ,
    \label{unitary_constraint}
\end{equation}
i.e., $\bm{L}$ is a subunitary matrix when losses are present in the system.
Hence, the action of the noisy LON on an $m$-mode coherent state $\ket{\bm{\gamma}}$ reads
\begin{equation}
\begin{split}
\mathcal{E}(\ketbra{\bm{\gamma}})=\Tr_{env}\lbrace{\mathcal{U} (\ketbra{\bm{\gamma}} \otimes \nu_{th}^{\otimes m}) \mathcal{U}^\dagger}\rbrace \, ,
\label{action_on_coherent}
\end{split}
\end{equation}
where $\mathcal{U}$ is the unitary operator associated with the larger $2m$-mode interferometer and the trace is taken over the environmental degrees of freedom.
\begin{figure}
    \centering
    \includegraphics[width=0.48\textwidth]{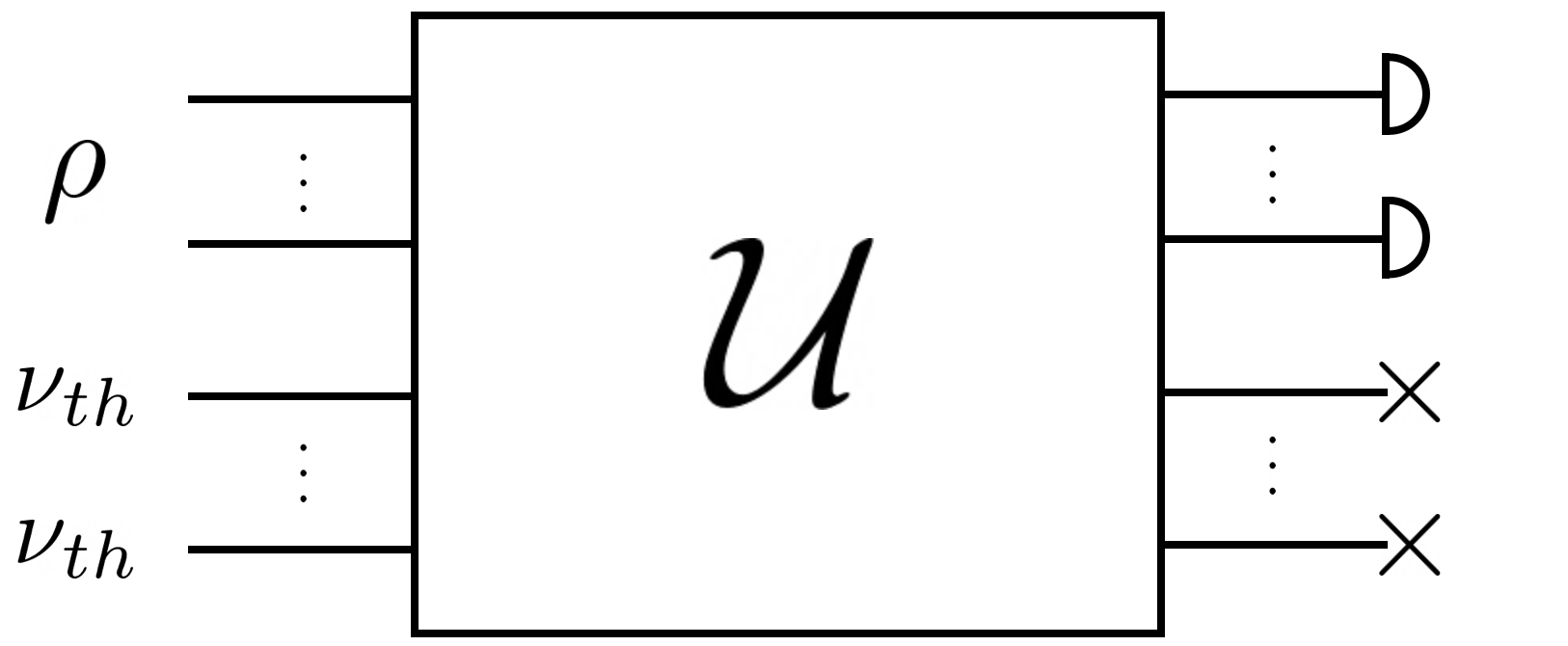}
    \caption{Schematics of the noise model used throughout this work, described by the CP map $\mathcal{E}$. The system's initial $m$-mode state $\rho$ interacts with the environment $-$ initialized in a thermal state $-$ through a loss-less $2m$-mode interferometer represented by a unitary operation $\mathcal{U}$. At the output ports, the system's modes are measured, while the crosses represent tracing over the environmental degrees of freedom.}
    \label{fig_thermal_noise}
\end{figure}
In Fig.~\eqref{fig_thermal_noise} we display a schematic representation of the noise model employed to describe the LON.
We can expand the $m$-mode thermal state over the coherent state basis, making use of its $P$-function representation 
\begin{equation}
    \nu_{th}^{\otimes m}=\int d^{2m}\bm{\beta} P_{th}(\bm{\beta})\ketbra{\bm{\beta}} \, ,
\end{equation}
with
\begin{equation}
    P_{th}(\bm{\beta})=\frac{2^m}{\pi^m(k-1)^m} e^{-\frac{2}{k-1}\bm{\beta}\bm{\beta}^\dagger} \, ,
    \label{P_function_thermal}
\end{equation}
where $k=2\overline{n}+1$.
We can thus write Eq.~\eqref{action_on_coherent} as
\begin{equation}
\begin{split}
    \mathcal{E}(\ketbra{\bm{\gamma}}) & =
    \int d^{2m}\bm{\beta} P_{th}(\bm{\beta})  \Tr_{env}\lbrace{\mathcal{U}\ketbra{\bm{\gamma},\bm{\beta}}\mathcal{U}^\dagger}\rbrace \, .
\end{split} 
\end{equation}
The action of the larger LON on a multi-mode coherent state is easily computed
\begin{equation}
\mathcal{U}\ket{\bm{\gamma},\bm{\beta}} = \ket{(\bm{\gamma},\bm{\beta})\bm{U}}  =
    \ket{\bm{\gamma} \bm{L}+\bm{ \beta} \bm{P},\bm{\gamma} \bm{N} +\bm{ \beta} \bm{Q}} \, ,
\end{equation}
thus leading to
\begin{equation}
\mathcal{E}(\ketbra{\bm{\gamma}}) =
    \int d^{2m}\bm{\beta} P_{th}(\bm{\beta})  \ketbra{\bm{\gamma} \bm{L} + \bm{\beta}  \bm{P}} \, .
    \label{action_on_coherent_expanded}
\end{equation}
By substituting this expression into Eq.~\eqref{expansion}, we can compute the trace appearing in Eq.~\eqref{transition_function}
\begin{equation}
        \Tr{\mathcal{E}(D^\dagger (\bm{\xi}))D(\bm{\zeta})}  =
    \pi^m \delta^{2m}(\bm{\xi}-\bm{\zeta}\bm{L}^\dagger) e^{\bm{\zeta} k(\bm{L}^\dagger \bm{L}-\mathbb{I}_m)\bm{\zeta}^\dagger /2}
    \, , 
    \label{trace}
\end{equation}
where we have used the identity
\begin{equation}
    \int \frac{d^{2m}\bm{\beta}}{\pi^{2m}} e^{\bm{\zeta}\bm{\beta}^\dagger-\bm{\beta}\bm{\zeta}^\dagger} = \delta^{2m}(\bm{\zeta}) \, ,
\end{equation}
as well as the unitarity constraint Eq.~\eqref{unitary_constraint} and standard multi-dimensional Gaussian integration.
We now plug Eq.~\eqref{trace} into Eq.~\eqref{transition_function} and finally obtain the transition function
\begin{equation}
    \begin{split}
        T_{\mathcal{E}}^{(\bm{s},\bm{t})}(\bm{\alpha},\bm{\beta}) & = \int \frac{d^{2m}\bm{\zeta}}{\pi^{2m}} e^{-\bm{\zeta}\bm{\Sigma}\bm{\zeta}^\dagger /2}  
        e^{(\bm{\beta}-\bm{\alpha} \bm{L})\bm{\zeta}^\dagger-\bm{\zeta}(\bm{\beta}^\dagger-\bm{L}^\dagger \bm{\alpha}^\dagger)}
        \\ & =
        \frac{2^m}{\pi^m\sqrt{\det{\bm{\Sigma}}}}e^{-2(\bm{\beta}-\bm{\alpha} \bm{L})\bm{\Sigma}^{-1}(\bm{\beta}^\dagger-\bm{L}^\dagger \bm{\alpha}^\dagger)}
        \, .
    \end{split}
\end{equation}
The latter is non-negative, well-behaved and has multi-variate Gaussian form \textit{iff} 
\begin{equation}
    \bm{\Sigma} = k(\mathbb{I}_m-\bm{L}^\dagger \bm{L})-\bm{s}+\bm{L}^\dagger \bm{t} \bm{L} \geq 0 \, .
    \label{positive_covariance_matrix}
\end{equation}
Furthermore, we can always find $\overline{\bm{t}},\overline{\bm{s}}\in\mathbb{R}^m$ such that the input state $\bm{t}-$PQD is non-negative for $\bm{t}\leq\overline{\bm{t}}$ and the $(-\bm{s})-$PQD associated with the quantum measurement is non-negative for $\bm{s}\geq\overline{\bm{s}}$.
Hence, the following inequality
\begin{equation}
    k(\mathbb{I}_m-\bm{L}^{\dagger}\bm{L})-\overline{\bm{s}}+\bm{L}^\dagger \overline{\bm{t}} \bm{L} \geq 0 
    \label{classicality_condition}
\end{equation}
constitutes our sufficient classicality condition for an efficient simulation of the sampling task described above to be feasible.
We emphasize once more that this condition is only sufficient. On the other hand, the modular nature of the formalism enables its wide applicability, our sole assumption being the noise model of the linear optical evolution given by Eq.~\eqref{action_on_coherent}. 
As expected, the results of Ref.~\cite{RahimiKeshari} are retrieved in the zero-temperature limit $k=1$.

\begin{figure}
    \centering
    \includegraphics[width=0.48\textwidth]{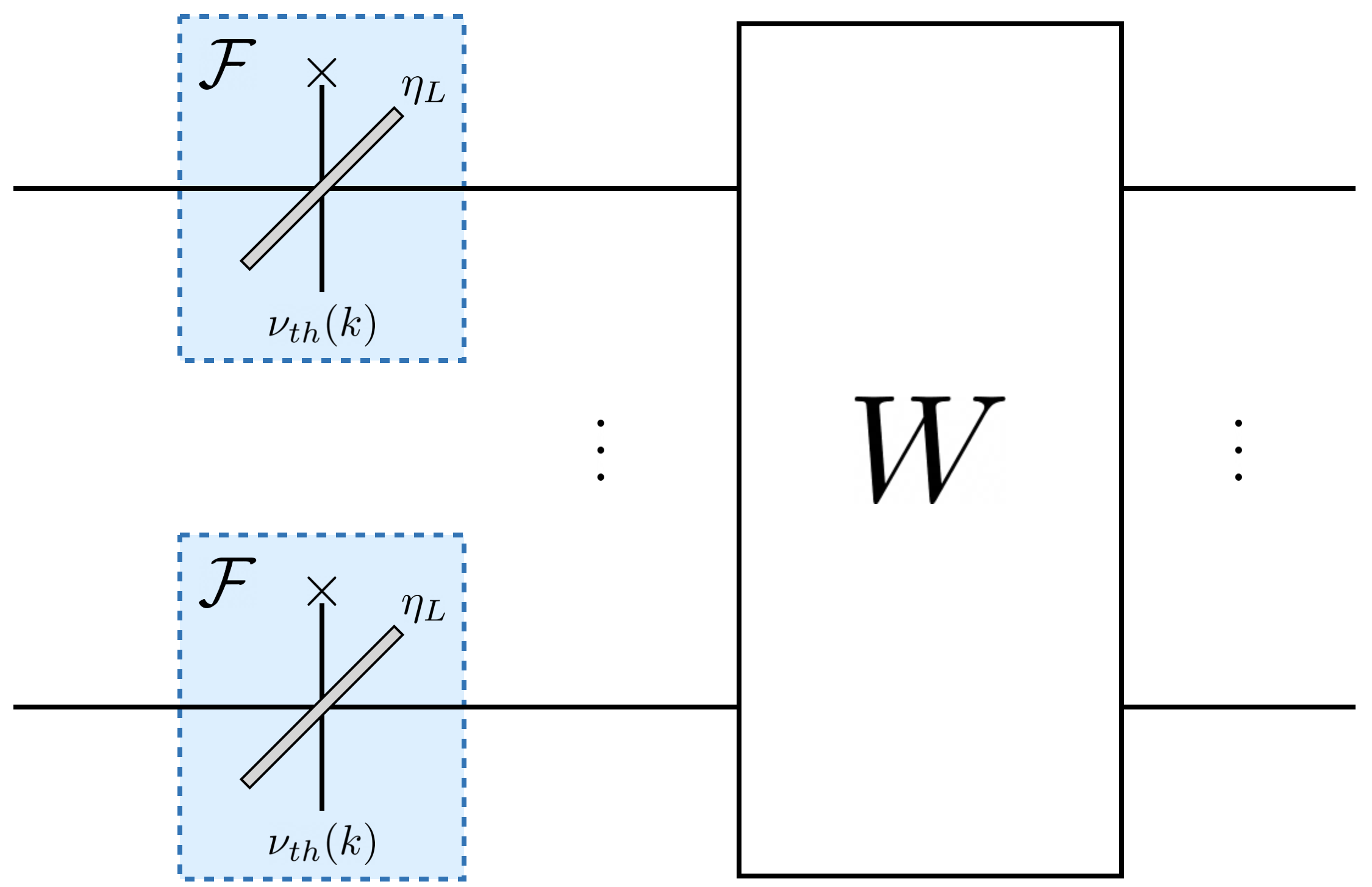}
    \caption{Decomposition of the CP map $\mathcal{E}$ describing the noisy linear optical evolution. Under the assumption of uniform losses we can absorb the latter and thermal noise effects into the initial state by means of the quantum channel $\mathcal{F}$, while retaining an ideal evolution described by the unitary matrix $\bm{W}$.}
    \label{fig_noise_decomposition}
\end{figure}

We can now apply Eq.~\eqref{classicality_condition} to a noisy GBS experiment. In particular, let us consider an initial state comprising of $m$ identical squeezed vacuum states $\bigotimes_{j=1}^m S(r)\ket{0}$, where 
\begin{equation}
    S(r)=e^{\frac{r}{2}(a^{\dagger 2}-a^{2})} \, 
    \label{squeezing}
\end{equation}
is the single-mode squeezing operator and $r>0$ is the squeezing parameter.
One can show that the $\bm{t}-$PQD of a generic $m$-mode Gaussian state $\rho$ reads
\begin{equation}
    W_{\rho}^{(\bm{t})}(\bm{\beta})=\frac{2^m}{\pi^m\sqrt{\det{{\bm{\sigma}}-{\tilde{\bm{t}}}}}}e^{-2(\bm{\beta}-\bm{\alpha})^\intercal({\bm{\sigma}}-{\tilde{\bm{t}}})^{-1}(\bm{\beta}-\bm{\alpha})} \, .
    \label{sPQD_Gaussian}
\end{equation}
Here, ${\bm{\sigma}}$ is the covariance matrix and $\bm{\alpha}$ is the displacement vector that fully characterizes $\rho$.
The conventions used are such that the covariance matrix of the single-mode thermal state Eq.~\eqref{thermal_state_def} is proportional to the identity matrix and reads $\bm{\sigma} = k\mathbb{I}_2 =(2\overline{n}+1)\mathbb{I}_2$.
Furthermore, ${\tilde{\bm{t}}}$ is a diagonal matrix defined as
\begin{equation}
    {\tilde{\bm{t}}}=\bigoplus_{j=1}^m t_j \mathbb{I}_2 \, .
\end{equation}
Consequently, the $\bm{t}-$PQD of a Gaussian state $\rho$ is non-negative \textit{iff}
\begin{equation}
    {\bm{\sigma}}-{\tilde{\bm{t}}}\geq0 \, .
    \label{gaussian_condition}
\end{equation}
If this condition is satisfied we will say that $\rho$ belongs to the set of $\bm{t}-$classical Gaussian states, which we denote with $\mathcal{C}^{(\bm{t})}_G$.
\\
For the input state $\bigotimes_{j=1}^m S(r)\ket{0}$, the above condition simplifies to 
\begin{equation}
    \bm{t}\leq e^{-2r} \mathbb{I}_m \equiv\overline{\bm{t}} \, .
    \label{t_bar_gbs}
\end{equation}
Let us also consider noisy threshold photo-detectors characterized by sub-unit efficiency $0\leq \eta_D \leq 1$ and by a dark count rate $0\leq p_D \leq 1$. The ``off'' and ``on'' elements of the POVM associated with this measurement respectively read 
\begin{equation}
    \Pi_0 = (1-p_D)\sum_{n=0}^\infty (1-\eta_D)^n \ketbra{n} \, ,
    \label{off_element}
\end{equation}
\begin{equation}
     \Pi_1=\mathcal{I}-\Pi_0 \, .
     \label{on_element}
\end{equation}
A close inspection of Eq.~\eqref{off_element} reveals that $\Pi_0$ is an unnormalized thermal state, hence one can analytically compute the $(-s)-$PQDs of both POVM elements using Eq.~\eqref{sPQD_Gaussian} and prove that they are non-negative for $s\geq 1-{2 p_D}/{\eta_D}$.
If we consider $m$ identical noisy threshold detectors as described above at the output ports of our LON, then the $(-\bm{s})-$PQD of the $m$-mode measurement is simply the product of the $(-s_j)-$PQD of the single-mode measurements, and it is clearly non-negative for 
\begin{equation}
    \bm{s}\geq\left(1-\frac{2 p_D}{\eta_D}\right)\mathbb{I}_m\equiv\overline{\bm{s}}\, .
    \label{s_bar_gbs}
\end{equation}
If one further assumes that losses are uniform across all possible paths across the network $-$ usually a good approximation for integrated setups $-$ then the linear transformation of the input modes is described by $\bm{L}=\sqrt{\eta_L} \bm{W}$, where $\bm{W}$ is a unitary matrix and $0\leq\eta_L\leq1$ denotes the transmission of the interferometer.
By substituting the threshold values $\overline{\bm{t}}$ and $\overline{\bm{s}}$ into Eq.~\eqref{classicality_condition} we obtain the classical simulability condition for the noisy GBS experiment  described above, i.e., 
\begin{equation}
    \frac{p_D}{\eta_D}\geq \frac{\eta_L}{2}(1-e^{-2r}) - \overline{n}(1-\eta_L)\, .
    \label{noisyGBS}
\end{equation}
The term $-\overline{n}(1-\eta_L)$ on the right-hand side (r.h.s.) of the inequality above represents a finite-temperature correction that accounts for thermal effects. 
Notice how the latter is always negative, meaning that thermal noise has the effect of reducing the detection noise needed for a classical simulation of the sampling task to be feasible, with respect to the zero-temperature scenario. 
Evidently, Eq.~\eqref{noisyGBS} is automatically satisfied whenever the r.h.s. becomes negative, i.e., if
\begin{equation}
    \overline{n}\geq \frac{\eta_L (1-e^{-2r})}{2(1-\eta_L)} \, . 
    \label{temperature_bound_GBS}
\end{equation}
We also remind the reader that the mean photon number of a thermal state is related to the environment's temperature $T$ via 
\begin{equation}
    \overline{n}=\frac{1}{e^{\frac{\hbar \omega}{k_{B} T}}-1} \, ,
\end{equation}
where $\omega$ is the mode's frequency and $k_B$ is the Boltzmann constant.
Therefore, Eq.~\eqref{temperature_bound_GBS} predicts the existence of a threshold temperature above which the sampling task becomes classically efficiently simulable even when ideal detectors are employed. 
This is somewhat expected, as it is well established that a noiseless boson sampling task with thermal input states leads to a classically simulable problem.
Consequently, we envision a transition in the computational complexity of the task as the environment's temperature increases.
\\
In the remaining part of this section we formalize these ideas and provide a physical interpretation of this phenomenon.
In Appendix \ref{appendix_absorb_noise} we show that the system's losses and thermal noise effects can be fully absorbed into the initial state, while retaining a unitary evolution via an effective ideal LON.
In particular, the noisy evolution given by $\mathcal{E}$ with $\bm{L}=\sqrt{\eta_L}\bm{W}$ is equivalent to a loss-less LON described by the unitary matrix $\bm{W}$, preceded by $m$ identical single-mode maps $\mathcal{F}$ that mix each input mode with a thermal state by means of a beam splitter with transmissivity equal to $\eta_L$, and finally taking the trace over the environmental degrees of freedom (See Fig.~\eqref{fig_noise_decomposition} for a schematic representation of the channel decomposition).
Hence, the action of the map $\mathcal{F}$ on a generic single-mode state $\rho$ reads
\begin{equation}
    \mathcal{F}(\rho) = \Tr_{env} \lbrace \mathcal{U}_{BS} (\rho \otimes \nu_{th}(k)) \mathcal{U}_{BS}^\dagger \rbrace \, .
    \label{single_mode_noise_map}
\end{equation}
Here, $\mathcal{U}_{BS}$ represents the beam splitter unitary operator acting on the system's mode and the corresponding ancillary environmental mode, and we have explicitly displayed the parameter $k$ that completely identifies the thermal state.
Focusing on GBS, each input mode of this loss-less LON is fed with $\mathcal{F}(S(r)\ketbra{0}S^\dagger(r))$, a Gaussian state whose covariance matrix reads $\text{diag}\lbrace a_+,a_-\rbrace$, with $a_{\pm}=\eta_L e^{\pm2r}+k(1-\eta_L)$.
As the covariance matrix of the vacuum state is simply the identity matrix, it is clear that quadrature squeezing vanishes if $a_-\geq 1$.
One then easily proves that this happens when condition Eq.~\eqref{temperature_bound_GBS} is satisfied.

As a result, the threshold temperature described in Eq.~\eqref{temperature_bound_GBS} can be physically interpreted as the temperature at which genuine quantum features of the input state completely disappear, so that efficient sampling on a classical machine becomes feasible regardless of the presence of noise in the detectors.
In the following section we extend this argument to a generic quantum optical experiment.

\section{Thermal classicalization}
\label{sec_thermal_classicalization}
Let us consider a generic $m$-mode input state $\rho$ undergoing noisy linear optical evolution via the quantum map $\mathcal{E}$ defined in the previous section, followed by a generic quantum measurement.
If we further assume that losses are uniform within the interferometer, then the classicality condition Eq.~\eqref{classicality_condition} may be recast as
\begin{equation}
    k\mathbb{I}_m \geq \frac{\overline{\bm{s}}-\eta_L \bm{W}^\dagger \overline{\bm{t}}\bm{W}}{1-\eta_L}
     \label{classical_simulability_condition_general} \, .
\end{equation}
The r.h.s. of this inequality (i.e., the threshold temperature) diverges at $\eta_L = 1$, which can be understood as the loss parameter also measures the ability of the LON to couple the system with the environment.
Eq.~\eqref{classical_simulability_condition_general} allows us to compute the temperature for an efficient classical simulation once the input state and the POVM have been specified.
Since $\overline{\bm{s}}\leq \mathbb{I}_m$ and $\overline{\bm{t}}\geq -\mathbb{I}_m$, it is clear that the most restrictive scenario is obtained by substituting $\overline{\bm{s}} = \mathbb{I}_m$ and $\overline{\bm{t}}= -\mathbb{I}_m$ into Eq.~\eqref{classical_simulability_condition_general}, resulting in
\begin{equation}
    k\geq\frac{1+\eta_L}{1-\eta_L}\, ,
\end{equation}
or equivalently
\begin{equation}
    \overline{n}\geq\frac{\eta_L}{1-\eta_L} \, .
    \label{simulate_everything_condition}
\end{equation}
To summarize, when the condition condition Eq.~\eqref{simulate_everything_condition} is satisfied, it enables the efficient simulation of a general noisy quantum optical experiment, our sole assumption pertaining to the model employed to describe the linear optical evolution.

We can also find a physical meaning to the threshold temperature given by Eq.~\eqref{simulate_everything_condition}.
We have already pointed out that the system's losses and thermal noise effects can be absorbed into the initial state by applying the map $\mathcal{F}$ to each input mode. Furthermore, we introduce the notation $\mathcal{F}_m \equiv \mathcal{F}^{\otimes m}$ for the corresponding $m-$mode map. 
\begin{equation}
\begin{split}
     & \mathcal{F}_m(\rho)  =\Tr_{env}\lbrace\mathcal{U}_{BS}^{\otimes m}(\rho\otimes\nu^{\otimes m}_{th}(k))\mathcal{U}_{BS}^{\dagger \otimes m}\rbrace
    \\ &  = \int d^{2m}\bm{\alpha} \, P_\rho (\bm{\alpha})\Tr_{env}\lbrace\mathcal{U}^{\otimes m}_{BS}(\ketbra{\bm{\alpha}}\otimes\nu^{\otimes m}_{th}(k))\mathcal{U}^{\dagger\otimes m }_{BS}\rbrace
     \\ & = \int d^{2m} \bm{\alpha} \, P_\rho (\bm{\alpha})\mathcal{F}_m(\ketbra{\bm{\alpha}})  
    \, ,
     \label{noisy_input_state}
\end{split}
\end{equation}
where we have exploited the Glauber $P-$function representation of $\rho$
\begin{equation}
    \rho = \int d^{2m}\bm{\alpha} \, P_\rho (\bm{\alpha}) \ketbra{\bm{\alpha}} \, . 
\end{equation}
Our objective is to compute the $P$-function of $\mathcal{F}_m(\rho)$, as it captures the non-classical properties of the noisy input state at study \cite{glauber,sudarshan}.
In particular, it is well known that having a well-behaved and non-negative $P-$function is a necessary and sufficient condition for a state to admit a classical description, i.e., it can be expressed as a statistical mixture of coherent states.
This property is also referred to as $P-$classicality.
As previously mentioned, the $P$-function is obtained by substituting $\bm{s}=\mathbb{I}_m$ into Eq.~\eqref{spqd_definition}, namely
\begin{equation}
    P_{\mathcal{F}_m(\rho)}(\bm{\beta}) =\int \frac{d^{2m}\bm{\xi}}{\pi^{2m}} \Tr\lbrace\mathcal{F}_m(\rho)D(\bm{\xi})\rbrace e^{\frac{\bm{\xi}\bm{\xi}^\dagger}{2}} e^{\bm{\beta}\bm{\xi}^\dagger-\bm{\xi}\bm{\beta}^\dagger} \, .
\end{equation}
We then substitute Eq.~\eqref{noisy_input_state} into the previous expression to obtain
\begin{equation}
   \begin{split}
    P_{\mathcal{F}_m(\rho)}(\bm{\beta}) & = \int \frac{d^{2m}\bm{\xi}}{\pi^{2m}} 
    \int d^{2m}\bm{\alpha} P_{\rho}(\bm{\alpha})
    \\ & \Tr\lbrace\mathcal{F}_m(\ketbra{\bm{\alpha}})D(\bm{\xi})\rbrace e^{\frac{\bm{\xi}\bm{\xi}^\dagger}{2} +\bm{\beta}\bm{\xi}^\dagger-\bm{\xi}\bm{\beta}^\dagger } \, .
\end{split} 
\end{equation}
The trace above may be computed using standard Gaussian calculation techniques, yielding
\begin{equation}
\Tr\lbrace\mathcal{F}_m(\ketbra{\bm{\alpha}})D(\bm{\xi})\rbrace = e^{-\frac{\lambda}{2}\bm{\xi}\bm{\xi}^\dagger+\sqrt{\eta_L}\bm{\xi}\bm{\alpha}^\dagger-\sqrt{\eta_L}\bm{\alpha}\bm{\xi}^\dagger} \, ,
\end{equation}
where $\lambda = k(1-\eta_L)+\eta_L$.
Putting everything together, we obtain
\begin{equation}
\begin{split}
    P_{\mathcal{F}_m(\rho)}(\bm{\beta}) & =\int\frac{d^{2m}\bm{\xi}}{\pi^{2m}}\int d^{2m}\bm{\alpha} \,  P_{\rho}(\bm{\alpha}) \, \\ & e^{\frac{1-\lambda}{2}\bm{\xi}\bm{\xi}^\dagger+\bm{\xi}(\sqrt{\eta_L}\bm{\alpha}^\dagger-\bm{\beta}^\dagger)-(\sqrt{\eta_L}\bm{\alpha}-\bm{\beta})\bm{\xi}^\dagger} \, .
\end{split}
\end{equation}
We then perform straightforward Gaussian integration and obtain
\begin{equation}
     P_{\mathcal{F}_m(\rho)}(\bm{\beta})=\frac{2^m}{\pi^m(\lambda-1)^m} \int d^{2m}\bm{\alpha} P_{\rho}(\bm{\alpha}) 
     e^{-\frac{2}{\lambda-1}\vert\bm{\beta} -\sqrt{\eta_L}\bm{\alpha}\vert^2} \, ,
\end{equation}
namely the convolution of $\rho$'s $P-$function and a Gaussian distribution. 
At the threshold temperature $-$ i.e., $k=\frac{1+\eta_L}{1-\eta_L}$ or equivalently $\lambda = 1+2\eta_L$ $-$ we have
\begin{equation}
 P_{\mathcal{F}_m(\rho)}(\bm{\beta})=\int d^{2m}\bm{\alpha} P_{\rho}(\bm{\alpha}) \frac{e^{-\vert\frac{\bm{\beta}}{\sqrt{\eta_L}}-\bm{\alpha}\vert^2}}{(\pi\eta_L)^m}
 \label{Pin_Pout_link}
\end{equation}
meaning that the $P$-function of the noiseless state $\rho$ and that of  $\mathcal{F}_m(\rho)$ are related by a Weierstrass transform (Gaussian filter). 
The latter has a smoothing effect on $P_\rho$ that removes its negativities and divergencies, resulting in a full suppression of the state's genuine quantum features, as we prove in the remainder of this section.
To this end, we recall that the $P-$function and the $Q-$function are related via the identity
\begin{equation}
    Q_{\rho}(\bm{\beta})= \frac{\bra{\bm{\beta}}\rho\ket{\bm{\beta}}}{\pi^m} = \int d^{2m} \bm{\alpha} \, P_\rho (\bm{\alpha}) \frac{e^{-\vert \bm{\beta} - \bm{\alpha}\vert^2}}{\pi^m} \, .
    \label{P_and_Q_link}
\end{equation}
Comparing Eq.~\eqref{Pin_Pout_link} and  Eq.~\eqref{P_and_Q_link} allows us to establish a connection between the $P-$function of $\rho$ and that of its noisy counterpart $\mathcal{F}_m (\rho)$
\begin{equation}
    P_{\mathcal{F}_m(\rho)}(\bm{\beta}) = \frac{1}{\eta_L^m}Q_\rho (\tfrac{\bm{\beta}}{\sqrt{\eta_L}}) \, .
\end{equation}
It is well known that any $Q-$function is positive definite and well behaved, thus proving our thesis: any input state $\rho$ becomes $P-$classical after interacting with a thermal state via a beam splitter with trasmissivity $\eta_L$, at the threshold temperature given by Eq.~(\ref{simulate_everything_condition}).
In particular, we find that the $P-$function of the noisy state $\mathcal{F}_m(\rho)$ is proportional to the $Q-$function of the input state $\rho$, properly rescaled.

It is worth noting the close connection between the temperature bounds derived here and the definition of non-classicality depth as introduced in Ref.~\cite{ncdepth}.

\section{Approximate classical simulation of Gaussian boson sampling under thermal noise}
\label{sec_approximate_sampling}

In Ref.~\cite{classical_simulation_noisy_GBS} the authors investigate the classical simulability of a noisy GBS experiment whose lossy linear optical evolution is given by Eq.~\eqref{zero_temperature_LON} and whose imperfect threshold detection is described by the POVM elements Eq.~\eqref{off_element} and Eq.~\eqref{on_element}.
As it will be clear in the following, their approach can account for \textit{approximate} sampling, thus overcoming the principal limitation of the formalism outlined in Section \ref{sec_sufficientconditions}.
However, this advancement is achieved at the expense of generality, limiting the applicability of this methodology to a noisy sampling task as described above.
In particular, it is showed that the latter may be efficiently simulated up to error $\varepsilon$ if the following (sufficient) condition is satisfied
\begin{equation}
    \sech{\left(\frac{1}{2}\Theta\left[ \ln{\left(\frac{1-2q_D}{\eta_L e^{-2r}+1-\eta_L}\right)} \right]\right)}>e^{-\frac{\varepsilon^2}{4m}} \, ,
    \label{quesada_bound}
\end{equation}
where $q_D=\frac{p_D}{\eta_D}$ and $\Theta(x)=\max{(x,0)}$ is the ramp function.
If the previous inequality does not admit a solution for any $0\leq\varepsilon\leq1$, then the classical simulation algorithm fails and we say that the Gaussian boson sampler has passed the non-classicality test.
Building on the approach of Ref.~\cite{classical_simulation_noisy_GBS}, we establish the following classical simulability condition that accounts for the influence of thermal noise
\begin{equation}
    \sech{\left( \frac{1}{2} \Theta \left[ \ln{\left(\frac{1-2q_D}{\eta_L e^{-2r}+k(1-\eta_L)}\right)}\right]\right)} \geq (1-\varepsilon^2)^{\frac{1}{m}} \, ,
    \label{updated_condition}
\end{equation}
where $k=2\overline{n}+1$ and $\overline{n}$ denotes the mean number of environmental thermal photons.
Furthermore, we show that at zero temperature Eq.~\eqref{updated_condition} constitutes a tighter bound than  Eq.~\eqref{quesada_bound}.
Additionally, note that it always exists $\tilde{\varepsilon}$ such that Eq.~\eqref{updated_condition} admits a solution for values of $\varepsilon$ such that $0 \leq \tilde{\varepsilon} \leq \varepsilon \leq 1$. Hence, we will say that the non-classicality test is passed if $\tilde{\varepsilon}$ is larger than some fixed (small) threshold.

As stated earlier, the assumption of uniform losses enables us to absorb all imperfections and thermal noise effects into the initial state state, while retaining an ideal linear optical evolution described by the unitary operator ${\mathcal{W}}$.
Each input port of this loss-less interferometer is fed with the single-mode Gaussian state
\begin{equation}
    \label{eq:noisy_state}
    \tau=\mathcal{F}(S(r)\!\ketbra{0}\!S^\dagger (0)) \, ,
\end{equation}
which has a diagonal covariance matrix that reads $\text{diag}\lbrace a_+,a_- \rbrace$ with $a_\pm = \eta_L e^{\pm 2r}+k(1-\eta_L)$.
The probability $p(\bm{n})$ of observing a specific measurement outcome $\bm{n}=(n_1,\dots,n_m)$ with $n_i \in \lbrace 0,1 \rbrace$ is thus given by 
\begin{equation}
    p(\bm{n}) = \Tr\lbrace {\mathcal{W}} \tau^{\otimes m} {\mathcal{W}}^\dagger \Pi_{\bm{n}} \rbrace \, ,
\end{equation}
where 
$\Pi_{\bm{n}} = \bigotimes_{i=1}^m \Pi_{n_i}$ is the POVM associated with $m-$mode noisy threshold photo-detection.
Let us now consider a related sampling problem, where the same loss-less LON is fed with $m$ identical $t-$classical Gaussian states $\tilde{\tau}$.
The resulting outcome probability distribution reads
\begin{equation}
    \tilde{p}(\bm{n}) = \Tr\lbrace {\mathcal{W}} \tilde{\tau}^{\otimes m} {\mathcal{W}}^\dagger \Pi_{\bm{n}} \rbrace \, .
\end{equation}
We can then use the classical simulability condition Eq.~\eqref{classicality_condition} $-$ setting $\bm{L}=\bm{W}$ unitary matrix and $\overline{\bm{s}}=(1-2q_D)\mathbb{I}_m$ $-$ to deduce that sampling (exactly) from the output state  $\mathcal{W}\tilde{\tau}^{\otimes n}\mathcal{W}^\dagger$ can be efficiently performed if $\tilde{\tau}$ is $t-$classical for $t\in [1-2q_D,1]$. 
The idea is that when the noisy input state $\tau$ is similar enough to $\tilde{\tau}\in\mathcal{C}^{(t)}_G$ for some $t\in [1-2q_D,1]$ the corresponding sampling problem can be efficiently simulated up to a small error.
In the remainder of this section, we formalize this intuition.
The total variational distance (TVD) between the two output probability distributions $p$ and $\tilde{p}$
\begin{equation}
    \frac{1}{2}\vert\vert p-\tilde{p} \vert\vert_1 = \frac{1}{2}\sum_{\bm{n}}\vert p(\bm{n})-\tilde{p}(\bm{n})\vert
\end{equation}
is upper bounded as follows 
\begin{equation}
\begin{split}
    \frac{1}{2}\vert\vert p-\tilde{p} \vert\vert_1 & \leq 
\frac{1}{2}\vert\vert  \mathcal{W}\tau^{\otimes m}\mathcal{W}^\dagger-\mathcal{W}\tilde{\tau}^{\otimes m}\mathcal{W}^\dagger\vert\vert_{tr}
     \\  & =\frac{1}{2}\vert\vert \tau^{\otimes m}-\tilde{\tau}^{\otimes m} \vert\vert_{tr} \\  & \leq \sqrt{1-F({\tau}^{\otimes m},\tilde{\tau}^{\otimes m})} = \sqrt{1-(F({\tau},\tilde{\tau}))^m} \, .
     \label{chain_of_inequalities}
\end{split}
\end{equation}
Here, $F(\rho,\tau) = (\Tr\lbrace\sqrt{\sqrt{\rho}\tau\sqrt{\rho}}\rbrace)^2$ denotes the quantum fidelity and 
\begin{equation}
    \vert\vert \rho-\tau \vert\vert_{tr} = \Tr{\sqrt{( \rho-\tau)^\dagger( \rho-\tau)}} 
\end{equation}
is the trace norm, which is invariant under unitary transformations acting on $\rho$ and $\tau$.
We emphasize that the bound on the TVD in Eq.~\eqref{chain_of_inequalities} is more stringent compared to the one presented in Ref.~\cite{classical_simulation_noisy_GBS}.
There, the authors exploited a generalization of the Pinsker's inequality to bound the total variational distance with the R\'enyi relative entropy. However, the latter is an unbounded quantity, which in turn leads to a looser bound that also explains why Eq.~\eqref{quesada_bound} does not always admit a solution.
On the other hand, in Eq.~\eqref{chain_of_inequalities} we bound the TVD with a quantity that is always less or equal than one.

As any $t-$classical state with $t\in[1-2q_D,1]$ leads to an efficiently simulable instance of GBS, we further minimize the TVD (i.e., maximize the fidelity) over all possible choices of $\tilde{\tau}$, namely
\begin{equation}
    \label{eq:inequality}
    \frac{1}{2}\vert\vert p-\tilde{p} \vert\vert_1 \leq \sqrt{1-(F_{\text{max}})^m}\leq \varepsilon \, ,
\end{equation}
where
\begin{equation}
    F_{\text{max}} = \max_{t\in[1-2q_D,1]}
    \max_{{\tilde{\tau}}\in\mathcal{C}_G^{(t)}}
    F(\tau,\tilde{\tau})\, .
\end{equation}
In Appendix \ref{appendix:derivation_approximate_bound} we show how to analytically carry out the optimization above, following the techniques outlined in Ref.~\cite{classical_simulation_noisy_GBS}.
Notice that, if $\tau$ is itself $t-$classical, then clearly fidelity is maximised (and equal to one) for $\tilde{\tau}=\tau$, and it is possible to efficiently simulate the sampling task exactly.
In general, we obtain
\begin{equation}
    F_{\text{max}} = \sech{\left( \frac{1}{2} \Theta \left[ \ln{\left(\frac{1-2q_D}{\eta_L e^{-2r}+k(1-\eta_L)}\right)}\right]\right)} \, ,
\end{equation}
which we then substitute into Eq.~\eqref{eq:inequality} to retrieve Eq.~\eqref{updated_condition}, i.e.
our final sufficient condition for classical simulability of noisy GBS at finite temperature .
As expected, we observe that in the zero temperature limit $k=1$ we obtain a bound that is more restrictive than that presented in Eq.~\eqref{quesada_bound}.
Lastly, notice that by fixing $\varepsilon = 0$ (i.e., sampling from the exact probability distribution of the noisy experiment), we retrieve Eq.~\eqref{noisyGBS}.

\section{Conclusions}
\label{sec_conclusions}
Using a phase space method based on the negativity of the relevant quasi-probability distributions, we have established a sufficient condition for the efficient classical simulation of generic quantum (linear) optical experiments affected by loss and  thermal noise. 
Our results show how finite temperature effects reduce the threshold of the system's imperfections that are sufficient for an efficient simulation of the experiment to be feasible. 
We then turned our attention to a GBS task employing threshold detectors, and provided a non-classicality condition in the form of an inequality involving the squeezing and noise parameters (photon loss, mean thermal photon number, detectors inefficiencies and dark count rates), that any potential candidate for an experimental demonstration of quantum advantage must satisfy.
Furthermore, we showed that there exist a 
threshold temperature at which any sampling experiment becomes efficiently simulable, even in the presence of ideal detectors.
We presented a physical interpretation of this phenomenon by establishing a connection with the vanishing of the genuine quantum features of the state.
We hope that this work inspires rigorous studies on the transition occurring in the  computational complexity of GBS subject to increasing levels of thermal noise.

\section{Acknowledgments}
G.B. is part of the AppQInfo MSCA ITN which received funding from the European Union’s Horizon 2020 research and innovation programme under the Marie Sklodowska-Curie grant agreement No 956071.
H.K. is supported by the KIAS Individual Grant No. CG085301 at Korea Institute for Advanced Study.
MSK acknowledges the KIST Open Research Programme, Samsung GRC programme and the KIAS visiting professorship.
The project was supported by the UK EPSRC through EP/Y004752/1 and EP/W032643/1.

\bibliography{biblio}

\appendix

\section{}
\label{appendix_absorb_noise}
In this section we show that the noisy linear optical evolution described by $\mathcal{E}$ in Eq.~\eqref{action_on_coherent} can be decomposed into a quantum channel $\mathcal{F}_m$ that accounts for all losses and thermal noise effects that may impact the system, followed by a unitary evolution corresponding to an ideal loss-less interferometer.
The key assumption underlying this simplification is that of uniform losses across all of the network's paths. 

In the main text we have showed that the action of a lossy LON at finite temperature on a multi-mode coherent state $\ket{\bm{\gamma}}$ $-$ a convenient choice of basis to characterize the quantum channel $-$ can be modeled as
\begin{equation}
\mathcal{E}(\ketbra{\bm{\gamma}}) =
    \int d^{2m}\bm{\beta} P_{th}(\bm{\beta})  \ketbra{\bm{\gamma} \bm{L} + \bm{\beta} \bm{P}} \, .
    \label{LON_on_coherent}
\end{equation}
Here $P_{th}(\bm{\beta})$ is the $P-$function of an $m-$mode thermal state, whose expression is given by Eq.~\eqref{P_function_thermal}, while $\bm{L}$ and $\bm{P}$ are two subunitary $m\times m$ matrices that satisfy Eq.~\eqref{unitary_constraint}.
At zero temperature, $P_{th}(\bm{\beta})$ reduces to the Dirac delta function $\delta^{2m}(\bm{\beta})$ and the effect of the LON on the state $\ket{\bm{\gamma}}$ simply reads 
\begin{equation}
    \mathcal{E}_0 (\ketbra{\bm{\gamma}}) = \ketbra{\bm{\gamma}\bm{L}} = \ketbra{\sqrt{\eta_L}\bm{\gamma}\bm{W}}\, ,
    \label{zero_temperature_LON}
\end{equation}
where we have used the uniform losses assumption $\bm{L}=\sqrt{\eta_L}\bm{W}$,  with $\bm{W}$ unitary matrix.
Eq.~\eqref{LON_on_coherent} can then be express as
\begin{equation}
     \mathcal{E}(\ketbra{\bm{\gamma}}) = \int d^{2m}\bm{\beta} \, P_{th}(\bm{\beta}) D(\bm{\beta}\bm{P})
    \mathcal{E}_0(\ketbra{\bm{\gamma}})
     D^\dagger(\bm{\beta} \bm{P}) \, .
    \label{almostgaussian}
\end{equation}
Upon performing the variable change $\bm{\beta}\rightarrow\bm{\beta}\bm{P}^{-1}$ we obtain
\begin{equation}
    \mathcal{E}(\ketbra{\bm{\gamma}}) = \int d^{2m}\bm{\beta}
    f(\bm{\beta})
    D(\bm{\beta})\mathcal{E}_0(\ketbra{\bm{\gamma}}) D^\dagger(\bm{\beta}) \, .
    \label{evolution_at_finite_temperature}
\end{equation}
Here
\begin{equation}
    f(\bm{\beta}) =  \frac{2^m}{\pi^m(k-1)^m(1-\eta_L)^m} e^{-\frac{2}{(k-1)(1-\eta_L)}\bm{\beta}\bm{\beta}^\dagger}
    \label{gaussian_displacing_distribution}
\end{equation}
is a multivariate Gaussian probability distribution
and we have used the fact that $(\bm{P}^{-1})^\dagger = (\bm{P}^{\dagger})^{-1}$ and $\bm{P}^\dagger \bm{P} = \mathbb{I}_m-\bm{L}^\dagger \bm{L} = (1-\eta_L)\mathbb{I}_m$.
Therefore, Eq.~\eqref{evolution_at_finite_temperature} indicates that the evolution of a coherent state through a lossy LON in the presence of thermal noise may be alternatively obtained by stochastically displacing the corresponding output state at zero temperature according to the Gaussian probability distribution Eq.~\eqref{gaussian_displacing_distribution}.
Furthermore, the linearity of $\mathcal{E}$ allows us to extend the aforementioned argument to an arbitrary input state $\rho$ and conclude that the evolved state at finite temperature $\mathcal{E}(\rho)$ can be obtained from the evolved state at zero temperature $\mathcal{E}_0(\rho)$ by applying the so called Gaussian noise/classical mixing.
The latter is a Gaussian CP map \cite{serafini2017quantum} whose action on the covariance matrix $\bm{\sigma}$ of a Gaussian  state reads \cite{ferraro2005gaussian}
\begin{equation}
    \bm{\sigma} \rightarrow \bm{\sigma} + (k-1)(1-\eta_L)\mathbb{I}_{2m} \, ,
    \label{gaussian_noise_on_covariance}
\end{equation}
while the vector of displacements is left unchanged.
Moreover, it is clear from Eq.~\eqref{zero_temperature_LON} that the map representing the evolution through the lossy LON at zero temperature $\mathcal{E}_0$ can be seen as $m$ single-mode pure loss channels with transmission $\eta_L$, followed by an ideal linear optical evolution described by the unitary matrix $W$.
We remind the reader that a pure loss channel may be implemented by combining the input mode with an ancilla mode initialized in the vacuum state via a beam splitter with trasmissivity $\eta_L$, followed by tracing out the ancillary degrees of freedom.

Additionally, we can then move the Gaussian noise before the loss-less LON, as it is easy to prove that the two maps commute. 
In fact, it suffices to recall that the evolution of a Gaussian state under a Gaussian unitary operation consists of updating its covariance matrix $\bm{\sigma}$ according to  
\begin{equation}
    \bm{\sigma}\rightarrow \bm{S}\bm{\sigma}\bm{S}^\intercal \, .
    \label{symplectic_evolution_on_covariance}
\end{equation}
Here $\bm{S}$ is the symplectic matrix that corresponds to the unitary evolution induced by the loss-less interferometer.
Furthermore, a passive LON can always be decomposed in terms of beam splitters and phase shifters exclusively, both of which have orthogonal symplectic matrices, which in turn implies that $\bm{S}$ is orthogonal as well, i.e., $\bm{SS}^\intercal=\mathbb{I}_{2m}$.
It is then straightforward to show that the two Gaussian CP maps Eq.~\eqref{gaussian_noise_on_covariance} and Eq.~\eqref{symplectic_evolution_on_covariance} commute.
Finally, standard calculations exploiting the Gaussian formalism reveal that the pure loss and Gaussian noise channels we previously described can be merged into a single CP map $\mathcal{F}_m$ whose action on a generic $m$-mode state $\rho$ reads 
\begin{equation}
    \mathcal{F}_m (\rho) = \Tr_{env}\lbrace\mathcal{U}_{BS}^{\otimes m} (\rho \otimes \nu_{th}^{\otimes m}(k)) \mathcal{U}_{BS}^{\dagger\otimes m}\rbrace \, .
\end{equation}
Here $\mathcal{U}_{BS}$ denotes the unitary operator that represents a beam splitter with transmissivity $\eta_L$, acting on the system's mode and the associated ancillary environmental mode.
The proof is concluded by observing that $\mathcal{F}_m$ can be further decomposed into $m$ identical single-mode maps $\mathcal{F}$ whose action is described by Eq.~\eqref{single_mode_noise_map} in the main text.

\section{}
\label{appendix:derivation_approximate_bound} 

In this section we summarize and exploit the techniques employed in Ref.~\cite{classical_simulation_noisy_GBS} to analytically optimize the quantum fidelity $F(\tau,\tilde{\tau})$ over $t\in[1-2q_D,1]$ and $\tilde{\tau}\in\mathcal{C}^{(t)}_G$, i.e.
\begin{equation}
    F_{\text{max}} = \max_{t\in[1-2q_D,1]}
    \max_{{\tilde{\tau}}\in\mathcal{C}_G^{(t)}}
    F(\tau,\tilde{\tau})\, .
\end{equation}
We remind the reader that $\tilde{\tau}$ is a $t-$classical Gaussian state, while $\tau$ is a noisy input (Gaussian) state whose covariance matrix reads $\bm{\sigma}_{\tau}=\text{diag}\lbrace a_+,a_- \rbrace$ with $a_\pm = \eta_L e^{\pm 2r}+k(1-\eta_L)$.
It is well known that any single-mode Gaussian state with zero displacement can be parametrized as a squeezed thermal state.
In particular, one easily proves that $\tau = S(r_\tau) \nu_{th}(\overline{n}_\tau) S^\dagger(r_\tau)$
with squeezing parameter and mean thermal photon number respectively given by  
\begin{equation}
\label{squeezing_parameter}
    r_{\tau} = \frac{1}{4} \ln{\left( \frac{a_+}{a_-} \right)} \, ,
\end{equation}
\begin{equation}
\label{mean_thermal_photon}
    \overline{n}_\tau = \frac{1}{2}(\sqrt{a_+ a_-}-1) \, .
\end{equation}
On the other hand, since the state $\tilde{\tau}= S(r_{\tilde{\tau}}) \nu_{th}(\overline{n}_{\tilde{\tau}}) S^\dagger(r_{\tilde{\tau}})$ is $t-$classical, its covariance matrix $\bm{\sigma}_{\tilde{\tau}}$ must satisfy Eq.~\eqref{gaussian_condition}, i.e. $\bm{\sigma}_{\tilde{\tau}}-t \mathbb{I}_2  \geq 0 $.
One then easily shows that this condition is equivalent to the following constraint
\begin{equation}
    \label{classicality_constraint}
    r_{\tilde{\tau}} \leq \frac{1}{2} \ln{\left(\frac{2 \overline{n}_{\tilde{\tau}}+1}{t}\right)} \equiv r_0(\tilde{\tau})\, .
\end{equation}
The fidelity between two single-mode Gaussian states $\tau$ and $\tilde{\tau}$ has a known analytical expression given by \cite{fidelity_Gaussian_states} 
\begin{equation}
    F(\tau,\tilde{\tau}) = \frac{1}{\sqrt{\Delta + \Lambda}-\sqrt{\Lambda}} \, ,
    \label{fidelity_GS}
\end{equation}
where
\begin{equation}
\begin{split}
    \Delta & = \frac{1}{4}\det{\bm{\sigma}_{\tau}+\bm{\sigma}_{\tilde{\tau}}} \\ 
    & = (\overline{n}_{\tau}-\overline{n}_{\tilde{\tau}})^2 + (2\overline{n}_{\tau}+1)(2\overline{n}_{\tilde{\tau}}+1)\cosh^2{(r_\tau - r_{\tilde{\tau}})} \, ,
\end{split}
\end{equation}
\begin{equation}
\begin{split}
        \Lambda & = \frac{1}{4}(\det{\bm{\sigma}_{\tau}}-1)(\det{\bm{\sigma}_{\tilde{\tau}}}-1) \\ 
        & = 4 \overline{n}_{\tau}(\overline{n}_{\tau} +1)\overline{n}_{\tilde{\tau}}(\overline{n}_{\tilde{\tau}}+1)\, .
\end{split}
\end{equation}
We first optimize $F(\tau,\tilde{\tau})$ over all possible $t-$classical states $\tilde{\tau}$, i.e. we find the point $(\overline{n}_{\tilde{\tau}}^*,r_{\tilde{\tau}}^*)$ that maximizes the fidelity at fixed $t$, subject to the constraint Eq.~\eqref{classicality_constraint}.
Clearly, if  $\tau$ is itself a  $t-$classical state $-$ i.e. $r_{\tau}\leq r_0(\tau)$ $-$ it follows that fidelity is maximised for $(\overline{n}_{\tilde{\tau}}^*,r_{\tilde{\tau}}^*) = (\overline{n}_\tau,r_\tau)$ and $F_{\text{max}} = 1$ for every value of $t$.
On the other hand, when $r_{\tau} > r_0(\tau)$ we find that the maximum fidelity at fixed $t$ reads $\sech\left( r_\tau - {1}/{2} \ln{\left({(2\overline{n}_\tau+1)}/{t}\right)}\right)$.
One then easily notices that the further optimization of this expression over $t\in[1-2q_D,1]$ is achieved by setting $t=1-2q_D$.
Finally, using Eq.~\eqref{squeezing_parameter} and  Eq.~\eqref{mean_thermal_photon}, and combining the two optimization regimes using the ramp function $\Theta$, after some algebra we obtain the desired result, namely
\begin{equation}
    F_{\text{max}} = \sech{\left( \frac{1}{2} \Theta \left[ \ln{\left(\frac{1-2q_D}{\eta_L e^{-2r}+k(1-\eta_L)}\right)}\right]\right)} \, .
\end{equation}
\end{document}